\documentclass[structabstract]{aa}
\usepackage[]{graphicx}
\usepackage{natbib}
\bibpunct{(}{)}{;}{a}{}{,} 
\usepackage{txfonts}
\graphicspath{{astrofigs/},{./},{figs/}}
\newcommand{\pdv}[2]{\frac{\partial #1}{\partial #2}}

\begin{document}

\title{Structures in compressible magnetoconvection
       and the nature of umbral dots}
       \author{C. Tian\inst{1} \and K. Petrovay\inst{2}}
\institute{Department of Physics,
           Gustaf H\"allstr\"omin katu 2a (PO Box 64),
          FI-00014 University of Helsinki, Finland
           \and
           E\"otv\"os University,
           Department of Astronomy,
           Budapest, Pf.~32, H-1518 Hungary}

\date{Received; accepted}

\abstract
{Structures seen in idealized numerical experiments on
compressible magnetoconvection in an imposed strong vertical
magnetic field show important differences from those detected 
in observations or realistic numerical
simulations of sunspot umbrae.}
{To elucidate the origin of these discrepancies,
we present a series of idealized 3D compressible
magnetoconvection experiments that differ from previous such
experiments in several details, bringing them closer to
realistic solar conditions.}
{An initially vertical magnetic field $B_0$ is imposed on a
time snapshot of fully developed solar-like turbulent
convection in a layer bounded by a stable layer from above. Upon
relaxation to a statistically steady state, the structure of the flow
field and magnetic field is examined.}
{Instead of the vigorous granular convection (GRC) well known to take
place in unmagnetized or weakly magnetized convection, for high values
of $B_0$ heat is transported by small-scale convection (SSC) in the form
of narrow, persistent convective columns consisting of slender upflows
accompanied by adjacent downflow patches, which are reminiscent of the
``convectons'' identified in earlier semianalytic models. For moderate
field strengths, flux separation (FXS) is observed: isolated field-free
inclusions of GRC are embedded in a strongly magnetized plasma with SSC.
Between the SSC and FXS regimes, a transitional regime (F/S) is
identified where convectons dynamically evolve into multiply segmented
granular inclusions and back.}
  {Our results agree in some aspects more closely with observed umbral
structures than earlier idealized models, because they do reproduce the
strong localized, patchy downflows immediately adjacent to the narrow
convective columns. Based on recent observations of umbral dots, we
suggest that in some cases the conditions in sunspot umbr{\ae} 
correspond to the newly identified F/S transitional regime.}


\authorrunning{C. Tian \and K. Petrovay}
\titlerunning{structures in magnetoconvection and the nature of umbral dots}
\maketitle


\section{Introduction}

The spectacular increase in the resolution of solar observations
experienced in the past decade has resulted in unprecedentedly detailed
images and movies of the small-scale structures seen in sunspots. This
in turn has rekindled interest in theoretical studies of
magnetoconvection in order to understand the origin of these structures.
These studies have resulted in some spectacular breakthroughs such as the first
ever numerical simulation of whole sunspots
(\citealp{Rempel+:spotsimu.pair}; \citeyear{Rempel+:spotsimu.1D};
\citealp{rempel2012}).
Despite these successes, however, there are still many questions to be
answered. They include the exact nature of umbral
substructures, such as umbral dots, light bridges and dark nuclei, and
their relation to structures seen in different numerical
magnetoconvection experiments.

High resolution imaging and spectropolarimetric observations with the
Swedish Solar Tower (\citealp{Sobotka+Hanslmeier:SST};
\citealp{Riethmuller+:SST}; \citealp{Sobotka+Puschmann:SST};
\citealp{Ortiz+:SST}), Hinode (\citealp{Kitai+:Hinode};
\citealp{Riethmuller+:Hinode}; \citealp{Bharti+:Hinode.lanes},
\citeyear{Bharti+:Hinode.downflows}; \citealp{Sobotka+Jurcak:Hinode})
and the Dunn Solar Telescope (\citealp{Bharti+:DST};
\citealp{Rimmele:DST}) have shown that umbral dots (UDs) have
characteristic sizes of $\sim 200\,$km and that they are mostly short-lived,
the majority having lifetimes of a few times ten minutes or less. Larger
UDs often split or coalesce, and in high resolution images, dark lanes are
often seen inside them. In the bright parts of UDs an upflow of $\sim
1\,$km/s has been unambiguously detected in many instances. Strong
localized downflow patches have been detected on the periphery of UDs.
This association with up- and downflows clearly demonstrates that UDs
are a manifestation of magnetoconvection in the strongly magnetized,
compressible umbral plasma. Kilcik et al. (\citeyear{kilcik}) 
have studied the morphological and statistical properties of bright UDs
by comparing the high resolution data from New Solar Telescope at the
Big Bear Solar Observatory and 3D magnetohydrodynamic (MHD) simulations. 
Based on good seeing,
Watanabe et al. (\citeyear{watanabe2012}) studied the temporal
evolution of different kinds of UDs in detail. These very recent studies show 
not only agreement but also discrepancies among observations made 
by different authors and numerical simulations, such as lifetime, 
size, shape, downflows, and correlations between different quantities.
We note that these studies looked at different sunspots, and the
simulated sunspots are also different from the observed ones. So far,
no clear picture has emerged either from observations or simulation
study regarding how the detailed properties of UDs  depend on the
properties of the sunspots (i.e. field strength,  overall size,
etc.). In this sense it is not clear to what degree  discrepancies
are simply due to the natural variation of sunspot properties and to
what degree there are true discrepancies between realistic
simulations and observed sunspots in terms of a model deficiency.

This paper is organized as follows. 
Section \ref{secobj} gives the context and objectives of this paper. 
Section \ref{secnum} describes the setup of our numerical model. Section
\ref{secrsl}  presents and discusses the results from the simulations, 
while Section \ref{seccon} concludes the paper.

\section{Context and objectives}
\label{secobj}

\subsection{Idealized models of magnetoconvection with an imposed vertical
field}

In the past decades, theoretical studies of magnetoconvection in a
strong, vertically oriented magnetic field have gone a long way,
starting from nonlinear extensions of the classic linear stability
analysis of \cite{Chandrasekhar:book}. These
studies have proceeded through the modelling of convecting
magnetized Boussinesq and anelastic fluids in two and three dimensions to
modelling 3D compressible magnetoconvection in a polytropically
stratified atmosphere (\citealp{Weiss+:3Dcompr},
\citeyear{Weiss+:fluxsep}). One important realization in these
studies was that the spatial separation, a mechanism that
separates strongly magnetized
non-convecting plasma from weakly magnetized convective flows,
allows for overturning convection to also take place in situations where
linear theory would predict convective stability or overstability.

As the field strength is decreased from a value high enough to
completely suppress motions, convection first sets in in the form of
narrow, needle-like vertical convective columns. The horizontal scale of
these structures is smaller than the natural scale of unmagnetized
convection in the same layer, so this stage is usually referred to as
``small-scale convection'' (SSC).  While these convective columns can
even exist as solitary ``convecton'' solutions
(\citealp{Blanchflower:convecton2D};
\citealp{Blanchflower+:convecton3D}; \citealp{Houghton+:convecton}),
they typically appear in large numbers.  The ``convectons'' of earlier
Boussinesq models (\citealp{Blanchflower:convecton2D};
\citealp{Blanchflower+:convecton3D}) are vertically elongated convective
cells, with upflows and adjacent downflows embedded in stagnant fluid.
In later compressible experiments (\citealp{Weiss+:3Dcompr};
\citealp{Houghton+:convecton}), however, SSC takes the form of isolated,
convective upflow plumes, their mass flux being compensated for by a slow,
uniform sinking motion in the surrounding magnetized fluid.
The plumes are persistent and immobile, though they pulsate irregularly
for a weaker imposed field .

With further reduction of the imposed magnetic field, a phenomenon known
as {\it flux separation} (FXS) takes place: large isolated patches of
unmagnetized, vigorously convecting fluid appear inside the strongly
magnetized component where SSC is still going on. The character of the
vigorous convection in field-free patches is similar to the normal
granular convection (GRC) in the unmagnetic case. Flux separation was
first detected in a simulation by \cite{Tao+:fluxsep}; however, it seems
to be a more fragile mode of magnetoconvective energy transport than the
SSC as in simulations without full compressibility or
with lower aspect ratios it is not present. Also, magnetoconvective
systems show a hysteresis-like behaviour where the presence or absence of
FXS is not uniquely determined by the physical parameters of
the system but also depend on its history (\citealp{Weiss+:fluxsep}).

For even lower field strengths, the strongly magnetized component
becomes confined to an intermittent network of channels separating the
vigorously convecting patches. Finally, with the further decrease in the
imposed field the topology changes, giving way to the well known pattern
of isolated magnetic flux concentrations arranged in a magnetic
network-like pattern between granules (GRC).

The sequence described above, characterized as a competition between two
modes of convection (columnal SSC  vs. vigorous GRC), suggests an 
interpretation of UDs as the photospheric manifestation of SSC. 
At the same time, 
there are no obvious examples of FXS in the solar photosphere, and the
short lifetimes of observed UDs stand in stark contrast to the
persistent character of the SSC columns described above.

\subsection{Numerical simulations of magnetoconvection in sunspot umbrae}

Models like those described above are usually called ``idealized'' because
they aim at a systematic study of increasingly general magnetoconvective
systems that are still simple enough to allow parameter studies and to
be understood in relatively simple physical terms. In consequence, they
do not aspire to faithfully reproducing all the physical conditions
relevant to the solar photosphere; instead, they consider convection in
a polytropically stratified perfect compressible fluid, with closed
boundaries and fixed diffusivities.

Another approach to the problem consists in attempting to simulate
conditions prevailing in the solar photosphere as realistically as
possible, including radiative transfer and a realistic equation of state
in a fully compressible equilibrium model
(\citealp{Moradi+:spotstruct.review}). An obvious drawback to this
approach is that, when going to the very limits of the available computing
power, it is left with too few resources for systematic parametric
studies. Nevertheless these ambitious models have been quite successful in e.g.
reproducing the observed penumbral fine structure. Structures readily
identified with the observed umbral dots are also seen in such
simulations, as first demonstrated by \cite{Schussler+Vogler}. In
contrast, with the SSC columns of idealized models,
these simulated UDs have finite lifetimes, and they show a
characteristic ``coffee bean'' structure, being crossed by a dark lane
along the long axis of their oval shape. Downflows are present near the
end points of the dark lane. The dark lane itself is due to an optical
depth effect due to the pileup of upflowing material below a magnetic
cusp above the UD. Similar UD-like structures are also seen in the
full-spot simulations of Rempel et al.\
(\citeyear{Rempel+:spotsimu.pair}, \citeyear{Rempel+:spotsimu.1D}).
Observationally, similar structures were found in  the cases studied by 
\cite{Bharti+:Hinode.lanes}, \cite{Rimmele:DST}, and \cite{Ortiz+:SST}. 
 
\subsection{Objectives}

Despite the impressive achievements of these state-of-the-art simulations, some
differences between the detailed properties of observed and simulated umbral
dots remain (cf.~\citealp{Bharti+:UDsimu}). Even more important differences
persist, however, between the nature of magnetoconvective structures in 
idealized models, on the one hand, and in observed or simulated sunspots, 
on the other.
These differences include the contrast between the observed localized patchy
downflows next to UDs vs.\ the lack thereof around SSC
columns in experiments and the persistent character of convective columns vs.\ the
limited lifetimes of UDs.

We cannot claim a satisfactory understanding of magnetoconvection in sunspots
until the origins of these differences are properly elucidated and a convincing
link between convective columns and UDs is forged. For this, there is a need to
bridge the gap between the idealized models and the realistic simulations. As a
first step in this process, in this paper we constructed an idealized
magnetoconvection model where, however, we relaxed certain constraints 
employed in earlier idealized models such as \cite{Weiss+:fluxsep} to 
bring the model closer to solar conditions. These were:
\begin{list}{--}{}
\item a two-layer model wherein the convectively unstable fluid is
	bounded by a stable layer from above;
\item open lower boundary conditions;
\item a subgrid closure scheme to account for the effects of
	small-scale turbulent transport;
\item overall stratification, while polytropic for simplicity, 
	to be reasonably similar to the actual solar case; 
\item a novel numerical scheme, the Bhatnagar-Gross-Krook-MHD 
	(BGK-MHD) scheme that is based
on gas-kinetic theory. We implemented a numerical solver for the induction
equation in addition to the BGK Navier-Stokes equations (BGK-NS) solver
developed by \cite{tian2007}. The BGK gas-kinetic scheme is particularly well
suited to the capture of near-discontinuities in the plasma flows.  This
admits a satisfactory stabilization of transition from initial perturbation to
fully developed turbulent convection.
\end{list}

In the simulations we first relax a convectively unstable layer of perfect gas
polytropically stratified under gravity. Then, a vertical uniform magnetic field
is imposed on a snapshot of the fully developed, statistically steady turbulent
convection. After a statistically steady state of magnetoconvection is attained,
we interpolate the results on a higher resolution mesh and resume the
calculations. Upon reaching steady state again, the properties of 
this system are examined. We construct a series of models with
different strengths of the imposed magnetic field.

 
\section{Numerical model}
\label{secnum}

In the current study, the 3D BGK-NS solver developed by  \cite{tian2007}
is coupled with an induction equation solver and  applied 
to solving the following resistive MHD equations
under a gravitational field:
\begin{eqnarray}
\label{ns1}
\partial\rho/\partial t &=& -\nabla \cdot \rho \vec{\varv},\\
\label{ns2}
\partial\rho \vec{\varv}/\partial t &=& -\nabla \cdot (\rho
\vec{\varv}\vec{\varv} -\vec{B}\vec{B}) -\nabla p_{\rm
tot}+\nabla\cdot\vec{\Sigma}+\rho\vec{g},\\
\label{ns3}
\partial E/\partial t &=&-\nabla\cdot[(E+p_{\rm tot})\vec{\varv}
-\vec{\varv}\cdot\vec{\Sigma}+\vec{F}_{\rm d}
-\vec{B}\vec{B}\cdot\vec{\varv}\nonumber\\
&& -\vec{B}\times\eta
(\nabla\times\vec{B}) ]+\rho \vec{\varv}\cdot\vec{g},\\
\partial\vec{B}/\partial t &=& -\nabla\cdot (\vec{\varv}\vec{B}
-\vec{B}\vec{\varv}-\eta\nabla\vec{B})
\end{eqnarray}
where $p_{\rm tot}=p_g+p_m$, $E=E_i+E_m+E_k$, $p_g$ is the gas
pressure, $p_m=\frac{1}{2}\vec{B}\cdot\vec{B}$ is the magnetic pressure,
$E_i$ is the internal energy, $E_m=\frac{1}{2}\vec{B}\cdot\vec{B}$ is
the magnetic energy, and $E_k=\frac{1}{2}\rho\varv^2$ is the kinetic energy.
Here, $\vec{g}$ is the gravitational acceleration, $F_d$ the diffusive heat
flux, $\vec{\Sigma}$ the viscous stress tensor, and $\eta$ the magnetic
resistivity. All the other symbols have their standard meanings.  

\subsection{Bou ndary conditions}

Side boundaries are periodic. The upper boundary is impenetrable, and
the density and internal energy are fixed:
\begin{equation}
	\vec{\varv} = \pdv\rho{t}=\pdv et=0.
  \label{bc1}
\end{equation}

The lower boundary is transmitting, so here only the horizontal
averages of density and internal energy are fixed at their initial
values (a slightly superadiabatic state), while the vertical 
derivatives of their fluctuations are set to zero:
\begin{equation}
	\frac{\partial \rho\vec{\varv}}{\partial z}=
\frac{\partial \rho'}{\partial z}=
\frac{\partial e'_i}{\partial z}= \pdv{\overline{\rho}}t=
 \pdv{\overline{e}}t=0.
  \label{bc2}
\end{equation}
Throughout this paper, overbar and prime denote the average and fluctuating
parts, i.e. for a variable $a$ we have $a=\overline a+a'$. Unless
explicitly stated otherwise, overline denotes a horizontal average.

The magnetic field is constrained to become vertical,
\begin{equation}
B_x=B_y=\frac{\partial B_z}{\partial z}=0,
\end{equation}
at the top boundary.
The lower boundary is transmitting for magnetic field
\begin{equation}
\frac{\partial B_x}{\partial z}=\frac{\partial B_y}{\partial z}
=\frac{\partial^2 B_z}{\partial z^2}=0.
\end{equation}
In both cases, we have $\nabla\cdot\vec{B}=0$ at the boundaries.

\subsection{Initial setup}

Initially, the system consists of two layers: an upper stable
radiative layer ($1>z>z_1$) and a lower unstable convective
 layer ($z_1>z>0$). The
upper layer is subadiabatically stratified, the lower layer
is slightly superadiabatic:
\begin{eqnarray}
T(z)=\left\{\begin{array}{c}
     T_0[1+Z_1(1+z)] \quad \mathrm{\ for\ } 1>z>z_1,\\
     T_1[1+Z_2(1+z)] \quad \mathrm{\ for\ } z_1>z>0,
     \end{array}\right.
\end{eqnarray}
\begin{eqnarray}
p(z)=\left\{\begin{array}{c}
     p_0(T/T_0)^{(m_1+1)} \quad \mathrm{\ for\ } 1>z>z_1,\\
     p_1(T/T_1)^{(m_2+1)} \quad \mathrm{\ for\ } z_1>z>0,
     \end{array}\right.
\end{eqnarray}
where $Z_1$ and $Z_2$ are the dimensionless temperature gradients of
the subadiabatic and superadiabatic layers, respectively; $m_1$ and $m_2$
are the polytropic indices for radiative and convective layer,
respectively; $T_0$, $p_0$ are the temperature and pressure at the top
lid; $T_1$, $p_1$ are the temperature and pressure at the interface
between the initial stable and unstable layers, i.e.,
\begin{eqnarray}
  &T_0=T(z=1) \quad & T_1=T(z_1), \\
  &p_0=p(z=1) \quad & p_1=p(z_1).
  \label{layers2}
\end{eqnarray}
In hydrostatic equilibrium we have
\begin{equation}
  (m_1+1)Z_1=(m_2+1)Z_2=|\vec{g}|.
  \label{m2g}
\end{equation}
The choice of $Z_1$ and $Z_2$ determines the temperature contrast between
top and bottom, by fitting to a standard stellar model, the
vertical extent of our computational domain. For this reason $Z$ is
occasionally called a \emph{depth parameter} in the literature. For $m_2$
we choose a value below the adiabatic value that reproduces the right
value of gravity acceleration according to equation~(\ref{m2g}). For
$Z_1$ a lower value is adopted to make the top layer subadiabatic, thus
convectively stable. Then $m_1$ can also be determined from
equation~(\ref{m2g}). The specific values of these parameters are listed
in Table~\ref{pars}.
 
\subsection{Treatment of radiation and subgrid scale motions}

Viscosity and diffusivity due to subgrid scale turbulent eddies are
calculated according to the Smagorinsky model. The dynamical viscosity is
given by
\begin{equation}
\label{sgs0}
\mu=\rho
(c_{\mu}\Delta)^2(2\vec{\sigma}:\vec{\sigma})^{1/2},
\end{equation} 
where $c_{\mu}$ is an adjustable constant, usually chosen from the range
0.1--0.2. The filter width $\Delta$ is taken to be the local resolution,
a colon stands for tensor contract, and $\vec{\sigma}=\partial_i
\varv_j+\partial_j \varv_i$. The turbulent heat transfer coefficient
$C_S$ is calculated as
\begin{equation}
 C_S=\mu/\delta,
\end{equation}
where the Prandtl number $\delta$ is  assumed to be constant.

For the current study, radiative transfer is considered in the diffusion
approximation. The total (radiative$+$turbulent) diffusive heat flux is
then
\begin{equation}
\vec{F}_d=-C_T\nabla T -C_S \nabla S,
\end{equation}
where $S=C_p(\ln{T}-\nabla_a \ln{p})$ is the specific entropy, $C_p$ 
the specific heat at constant pressure, and $\nabla_a$ the adiabatic
gradient. In the stable layer, $C_T$ is set so that radiation carries
out the input energy flux. In the convection zone, $C_T$ is very close
to zero. As $C_S=\mu/\delta$ represents turbulent diffusion, it is set to
zero in the stable region.

\begin{table}
 \begin{center}
 \caption{Dimensional analysis.}
 \label{scls}
 \begin{tabular}{@{}lcr}
  \hline
  \hline
  Scale & value & unit  \\
  \hline
  $\rho_{\rm{scl}}$  & $8.11\times 10^{-8}$ & $\rm{g/cm^3}$\\
  $p_{\rm{scl}}$  & $2.95\times 10^4$ & $\rm{dyn/cm^2}$\\
  $T_{\rm{scl}}$  & $5063.52          $ & $\rm{K}$ \\
  $l_{\rm{scl}}$  & $1.72             $ & $\rm{Mm}$ \\
  $t_{\rm{scl}}$  & $285.30           $ & $\rm{s}$ \\
  $\varv_{\rm{scl}}$  & $ 6.03        $ & $\rm{km/s}$ \\
  $B_{\rm{scl}}$  & $ 2911.55          $ & $\rm{G}$ \\
  $F_{\rm{scl}}$  & $6.34\times 10^{11}$ & $\rm{erg/s/cm^2}$ \\
  \hline
 \end{tabular}
\end{center}

\end{table}

\subsection{Dimensional analysis and model parameters}

\begin{figure} 
\centering
\includegraphics[width=\columnwidth]{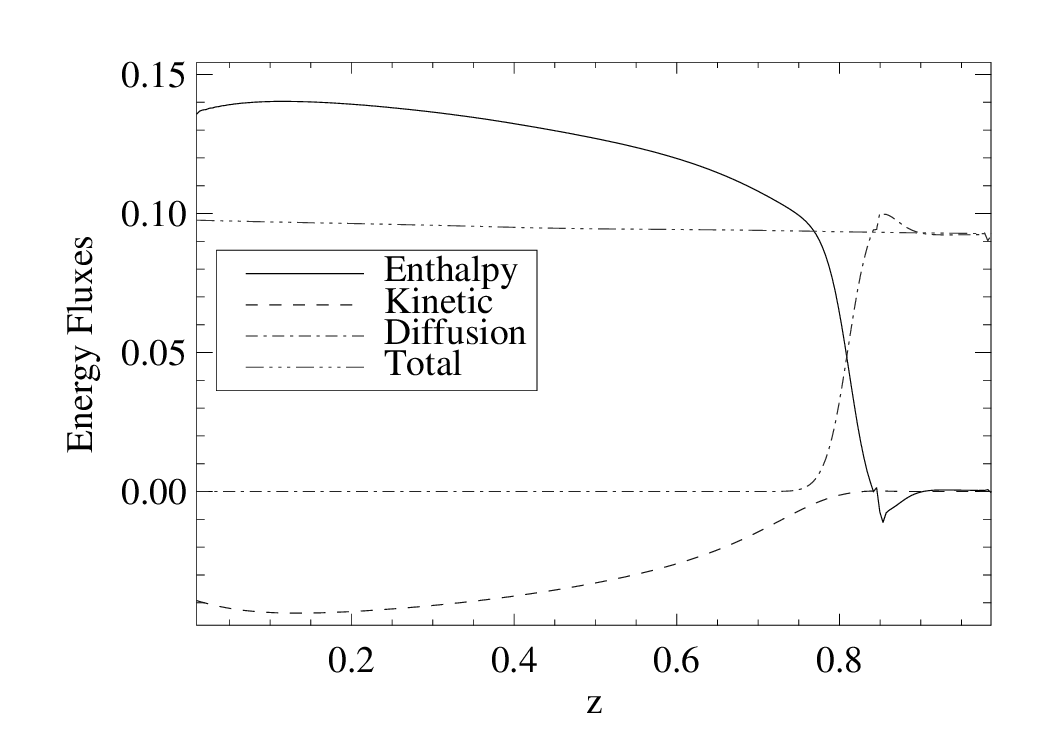}
\caption{
Statistical properties of steady unmagnetized convection, used as
initial condition for magnetoconvection. Shown are the  averaged
energy fluxes, in units of $6.34\times 10^{11} \rm{erg/s/cm^2}$ . 
Averages are taken in time and in the horizontal plane.  $z$ 
is the height from bottom, in units of $1.72 \rm{Mm}$.}
\label{figanls}
\end{figure}

\begin{figure}
\centering
\includegraphics[width=\columnwidth]{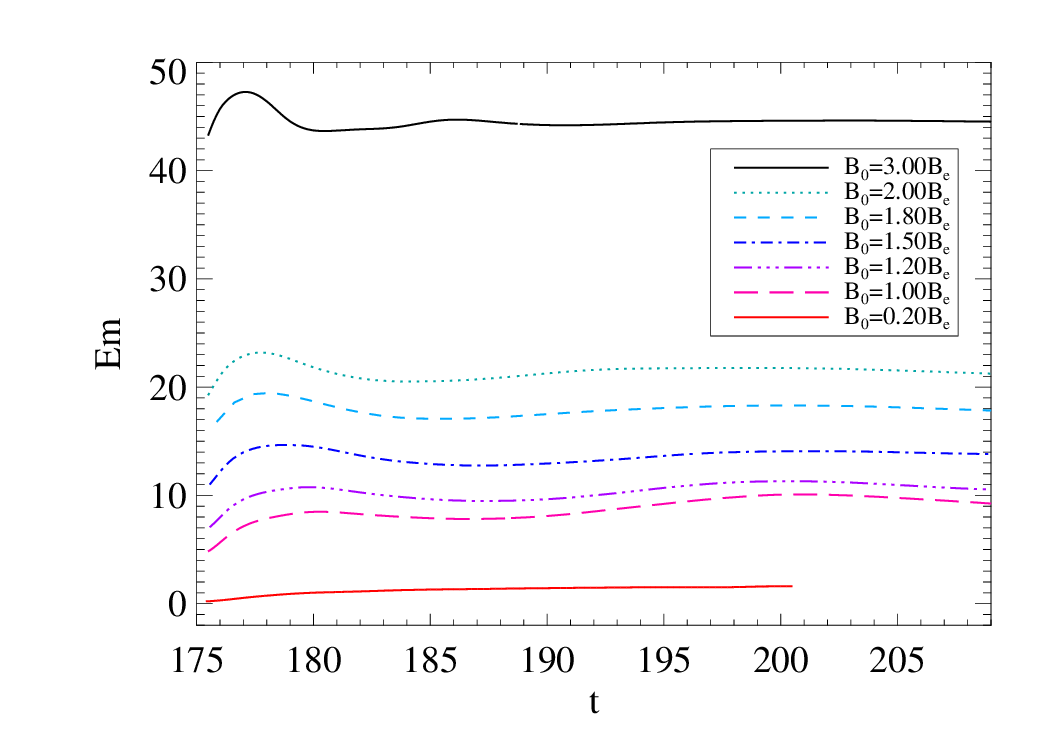}
\caption{Histories of magnetic energy for various runs, as indicated by
the labels.}
\label{figthme}
\end{figure}

We non-dimensionalize all variables by choosing the following basic
scales. The scale of the length, $l_{\rm scl}$, is the depth of the
computational domain. For the density, gas pressure, and temperature,
their respective values at  the top boundary are chosen as the basic
scales, i.e.,  $\rho_{\rm scl}$, $p_{\rm scl}$, and $T_{\rm scl}$. The
scales for the rest of the quantities can be deduced according to
dimensional analysis. For instance, the scale of velocity and
time are
\begin{equation}
  \varv_{\rm scl}=(p_{\rm scl}/\rho_{\rm scl})^{1/2} \qquad
  t_{\rm scl}=l_{\rm scl}/\varv_{\rm scl},
  \label{vtbscl}
\end{equation}
respectively. (Note that $\varv_{\rm scl}$ is the isothermal sound speed
at the top.) In the case of magnetic field strength, a different
definition is used in order to give it a more
physical meaning (of thermal equipartition field strength at the top
of the convective layer):
$B_{\rm scl}=2\pi^{1/2} p_{\ast}^{1/2}$, 
$p_\ast$ is the pressure at the top of the convection zone.
\begin{table} 
 \begin{center}
 \caption{Summary of numerical parameters.}

 \label{pars}
 \begin{tabular}{@{}lcc}
  \hline
  \hline
  Parameter & Definition & Value \\
  \hline
  Grid size     &$N_x\times N_y\times N_z$ & $256\times 256\times 256$  \\
  Aspect ratio  &$l_x : l_y:  l_z$ & $ 6: 6: 1$  \\
  CFL number    &$\Delta x/(\varv+c_s)/\Delta t$ & $ 0.3$  \\
  Depth of conductive zone    & in PSHs            & $1.87$  \\
  Depth of transition zone    & in PSHs            & $1.26$  \\
  Depth of convective zone    & in PSHs            & $3.61$  \\
  SGS Prandtl number & $\delta=\mu/\kappa$           & $1/3$  \\
  Deardorff number & $c_{\mu}$ in Eq.~(\ref{sgs0})   & $0.2$  \\
  Ratio of specific heat            & $\gamma$        & $5/3$  \\
  polytropic index        & $m_1$        & $2.728$  \\
  polytropic index        & $m_2$        & $1.485$  \\
  Depth parameter         & $Z_1$        & $2$  \\
  Depth parameter         & $Z_2$        & $3$  \\
  Initial convective top   & $z_1$        & $0.9$  \\
  \hline
 \end{tabular}
\end{center}

\end{table}

\begin{table}
 \begin{center}
 \caption{Imposed magnetic field strength in the numerical runs.}

 \label{runs}
 \begin{tabular}{@{}lc}
  \hline
  \hline
  Identifier & $B_0/B_e$ \\
  \hline
  A  &$3.00$  \\
  B  &$2.00$  \\
  C  &$1.80$  \\
  D  &$1.65$ \\
  E  &$1.50$  \\
  F  &$1.20$  \\
  G  &$1.00$  \\
  H  &$0.20$  \\
  \hline
 \end{tabular}
\end{center}
\end{table}

Of particular importance is the scale of the input energy flux,
\begin{equation}
  F_{\rm scl}=p_{\ast}{\varv_{\ast}},
  \label{fscl}
\end{equation}
where $\varv_{\ast}$ is the sound speed at the convective top.
As the solar radiation flux is fixed in physical units, its value in
non-dimensional units will depend on the value of $F_{\rm scl}$, i.e.\ on
the position of the upper boundary of our convection zone 
inside the solar convective envelope. Thus, by specifying the input flux, 
we can approximately control the radial location of our numerical computational
domain within the standard solar model.

In summary, the input energy flux, the aspect ratio and the depth of the
domain determine the ``geographic'' parameters of the model, i.e., size and
location. The scales of density, gas pressure, and temperature determine
the state of matter. To perform the dimensional analysis, we
need a standard solar model. In the current study, we adopt the
combined model calculated by \cite{Guenther1992} (Table 3B therein).

In our thermally relaxed unmagnetized convection model, the extent of
the top conductive layer ($z>0.9$) is $\sim 1.87 $PSHs (pressure scale 
heights) and the fully convective zone ($z<0.78$) $\sim 3.61$ PSHs. 
Between them, there is a transition layer of
around $\sim 1.26$ PSHs, including an upwardly overshooting zone.  
The solar-type granulation forms at around $z=0.8$, the lower 
part of the transition zone. We 
define $z=0.8$ as the effective top boundary of the convection zone. Thus the
upper conductive zone can be regarded as an artificial layer. 
The total energy flux
transported by the system is $\sim 0.1$, implying that the top of our convective
layer is  at a depth of $\sim 520\,$km from the solar surface (from unit
continuum optical depth). This distance roughly corresponds to the depth of the
Wilson depression. In such a case, the top of our box is already above the solar
photosphere, approaching the height of the temperature minimum. 

The scales of our models are summarized in Table~\ref{scls}.
Owing to the rapid change of pressure near the solar
surface, different standard solar models yield quite different
surface pressures, so our current dimensional analysis is
preliminary, especially the magnetic strength, which is determined
directly by the unit of gas pressure. 

For magnetoconvection it is in fact more meaningful to express magnetic
field strength in terms of the {\em turbulent equipartition field}
rather than in Alfv\'enic units (or thermal equipartition field).  The
turbulent equipartition magnetic field $B_e$ is defined by
$\rho\varv^2/2=B_e^2/8\pi$. We evaluate $B_e$ at  $z=0.8$,
the effective top of convection zone. We find
$B_e=0.51B_{\rm{scl}}=1484.89 \,$G. Table~\ref{runs} presents the strength
of the imposed vertical magnetic field in the individual runs in units
of $B_e$.

\begin{figure*}
\centering
\includegraphics[scale=0.85]{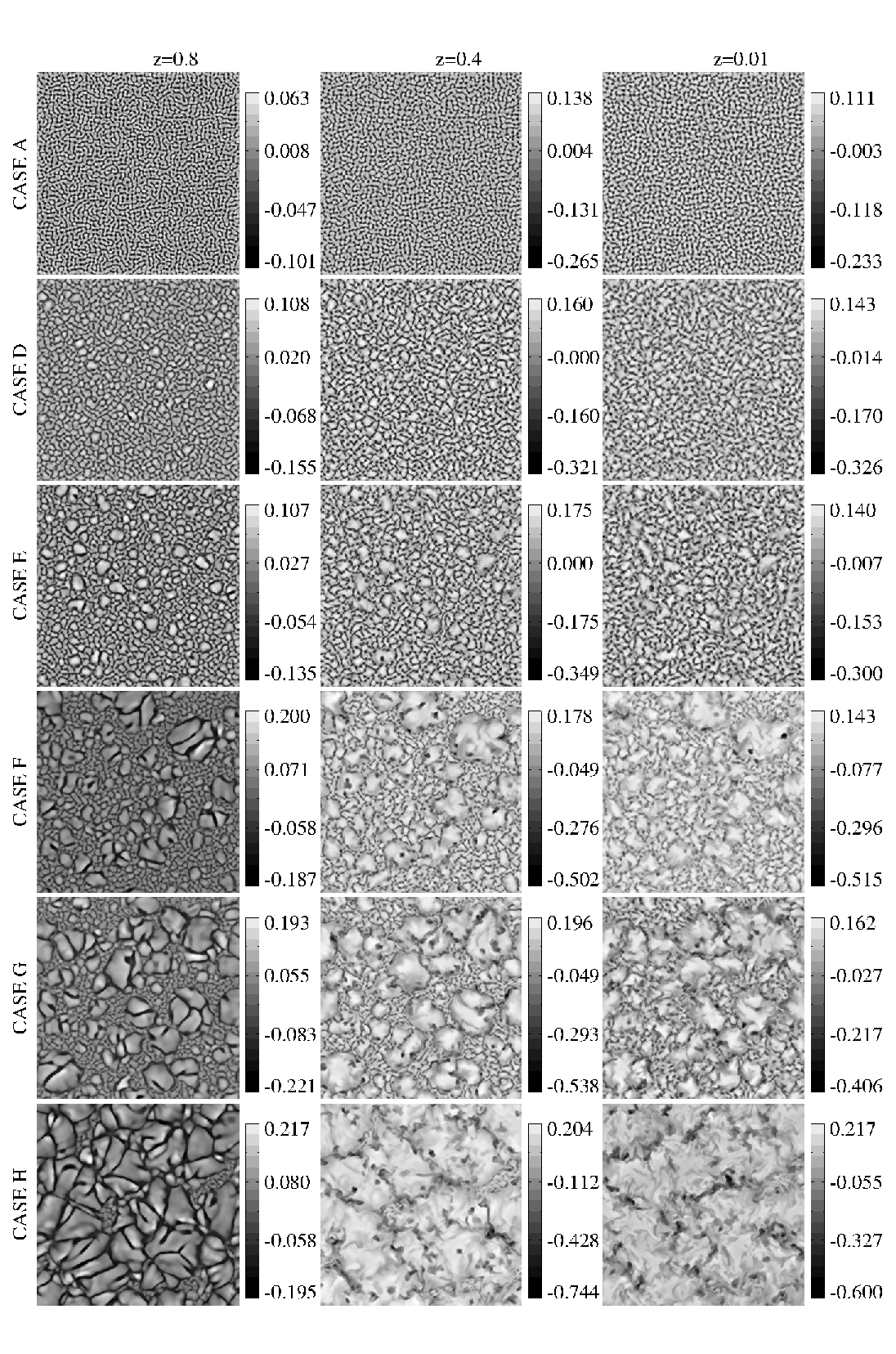}
\caption{Dopplergrams at various depths for selected cases. Corresponding
magnetic fields are shown in Fig.~\ref{figcop2}. Velocities are 
given in units of $6.03 \rm{km/s}$.}
\label{figcop}
\end{figure*}

\begin{figure*}
\centering
\includegraphics[scale=0.85]{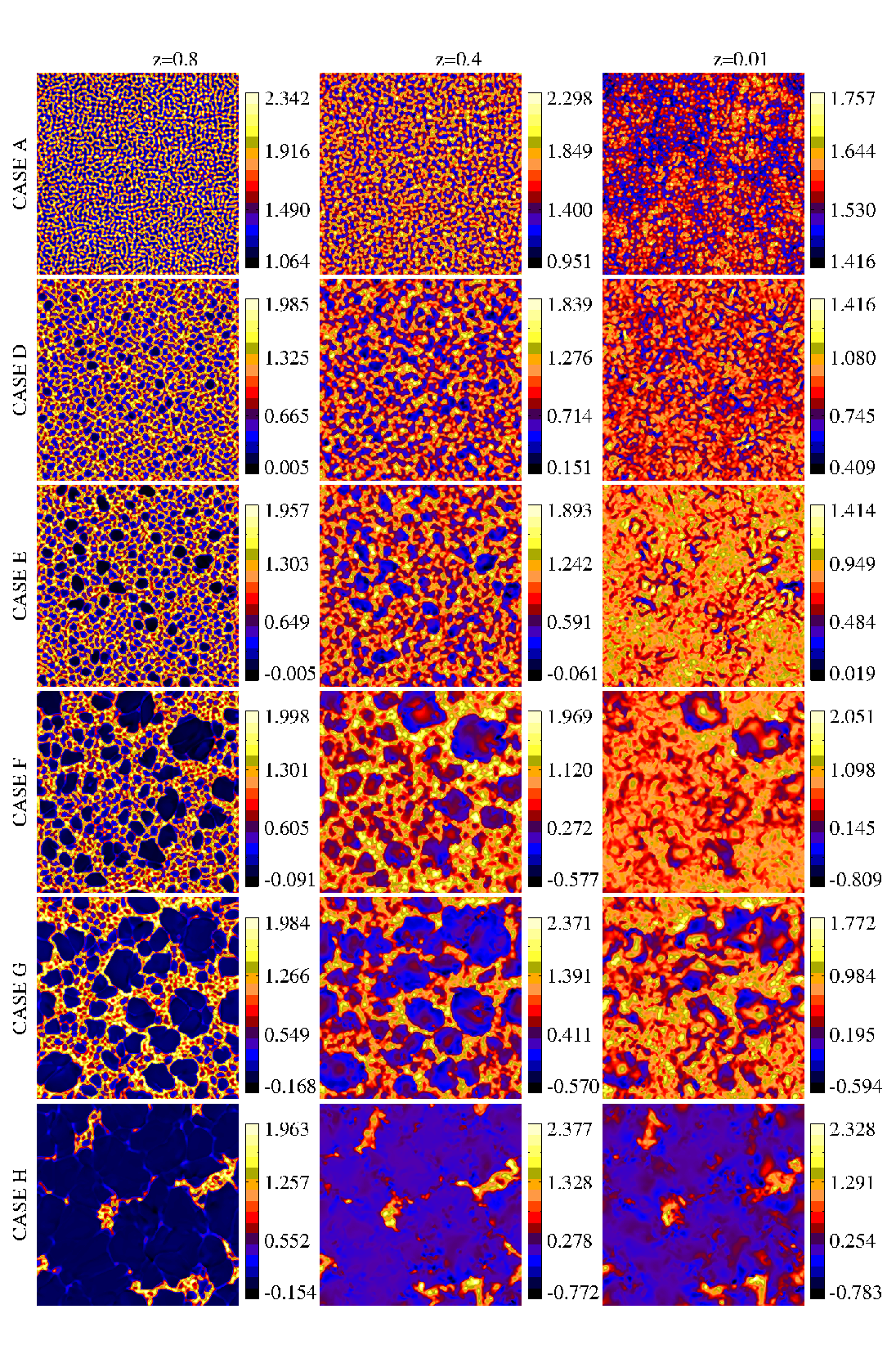}
\caption{Contour plots of vertical magnetic fields at various depth
for selected cases. Corresponding vertical flow fields are shown in
Fig.~\ref{figcop}. Magnetic fields are given in units of 
$2911.55 \rm{G}$. }
\label{figcop2}
\end{figure*}

\section{Numerical results and discussion}
\label{secrsl}

Upon starting the calculation with no magnetic 
field and after long term
relaxation, convection approaches a nearly steady state where the
temporally and horizontally averaged total energy flux is nearly
constant. The various contributions to the energy flux are shown in 
Fig.~\ref{figanls}. The strong downwardly directed
kinetic energy flux is a well known hallmark of steady compressible
convection, as expected from convection theory (\citealp{Unno+Kondo89};
\citealp{Petrovay90}) and as is also consistent with numerical simulations
(\citealp{Chan+Sofia86}). From the constancy of the total energy flux, it
is clear that the unmagnetized convection is completely relaxed.

Next, we superpose a vertical uniform magnetic field on an
instantaneous snapshot of the relaxed unmagnetized convection.
Figure~\ref{figthme} shows the time history of magnetic energy for
the cases listed in Table~\ref{runs}. It is apparent that after a
short while, the magnetic energy contained in the system becomes
statistically steady. For the cases with  weak initial 
fields, the magnetic fields are amplified by flow motions 
and eventually reach a nearly steady value. For the strong 
initial field cases, the magnetic energy
grows up to a maximum and then decreases to a nearly constant value.

\subsection{Flow morphology}

The effects of a vertical magnetic field imposed on the granulation
pattern are summarized in Figs.~\ref{figcop} and ~\ref{figcop2}.
As revealed by previous studies, e.g., \cite{Weiss+:3Dcompr} and
\cite{Stein+:magconv}, magnetic fields are intermittent for low 
field strengths, concentrated into flux tubes confined to the narrow lanes
of the intergranular network (case H). The strongest concentration
occurs at the junctions of the intergranular lanes of larger
granules in the deeper layer, forming pore-like structures. This 
is because in such cases, the strong magnetic fields are generated 
by amplification of flow motions. The downstream in the  
intergranular network of deeper bigger granules has a very long 
extent and persists for a long time, so it can continuously 
amplify the magnetic fields.

\begin{figure*}
\centering
\includegraphics{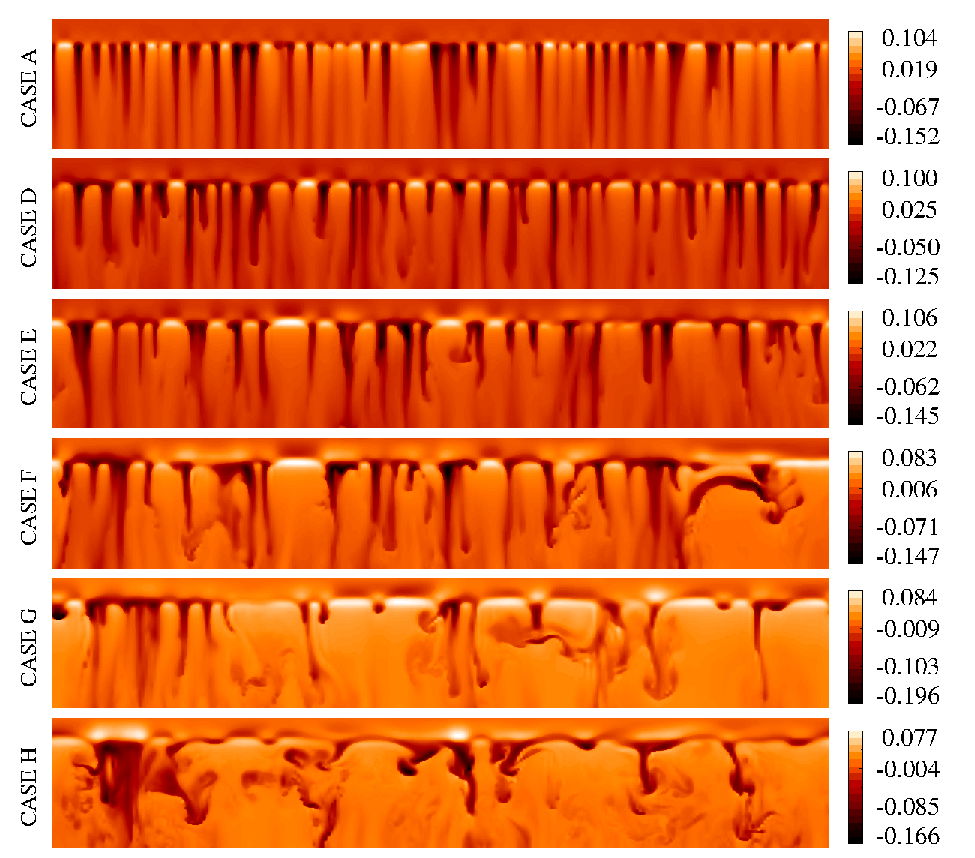}
\caption{Vertical cuts of the computational box 
along $x=3$ at an arbitrary instant. The colour scale 
represents temperature fluctuation
from its horizontal mean, i.e., $T'=T-\overline T$, in units 
of $5063.52 \rm{K}$.  Rows correspond to different runs as 
labeled on the left.}
\label{figcosvx}
\end{figure*}

\begin{figure*}
\centering
\includegraphics{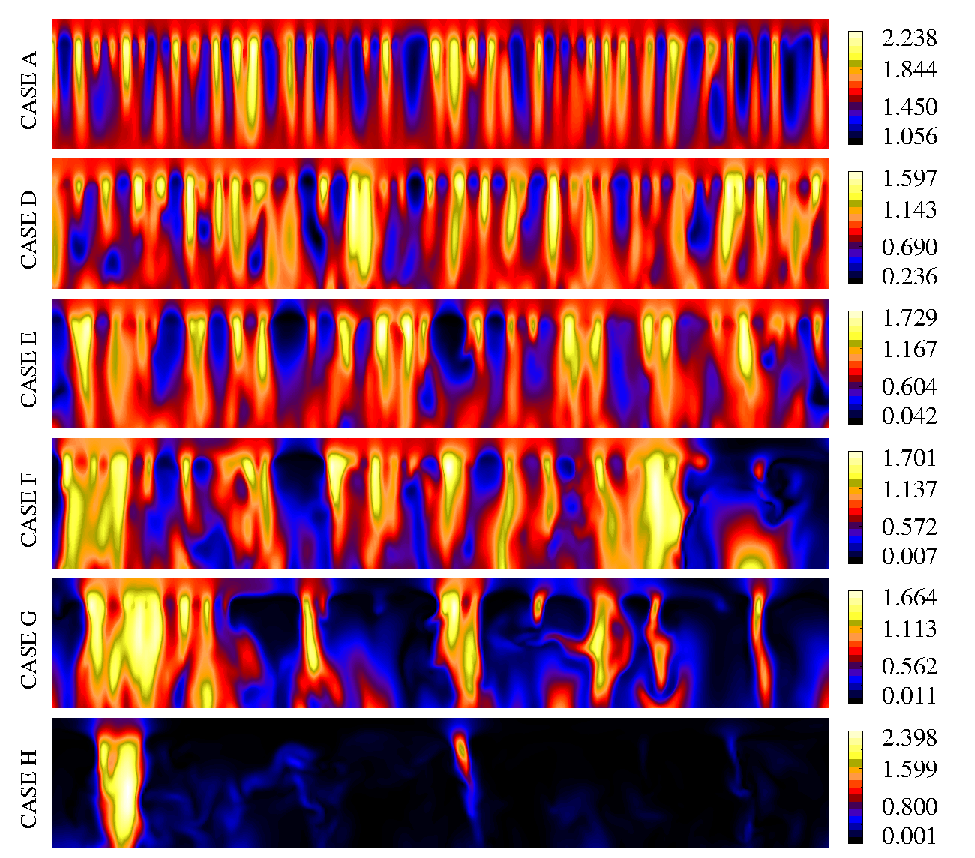}
\caption{Magnetic field strength $|B|$ corresponding to the cuts in Fig.~\ref{figcosvx}. The unit is $2911.55 \rm{G}$}
\label{figcosmx}
\end{figure*}

In the opposite limit of a very strong imposed magnetic field, convective
flows are significantly suppressed (cases A and B). The remaining
flows  and magnetic fields show a brainpattern configuration. At first
glance, it looks like a numerical effect. Since this kind of structure
is quite steady, we cut a piece of the computational domain and refined the
grids, then resumed the calculations. We repeated this procedure until
the horizontal resolution was improved by a factor of 8 in each
direction.  Numerical tests show that the brainpattern is an unresolved
weak convection pattern on a scale even smaller than the size of SSC. 
After doubling the grid points per unit, this kind of structure is already
resolved. Their shape persists when the resolution is increased further.
These high resolution results indicate they are indeed a very weak,
non-efficient, laminar, steady tiny scale convection pattern. The
fine unresolved steady convection is evident in the vertical cuts in the
upper panels of Figs. \ref{figcosvx} and \ref{figcosmx}. We expect
that implementing more realistic radiation transfer and boundary
conditions would suppress this fine structure, making it
undetectable in intensity maps and dopplergrams.

Case C acts like a critical situation, where a few small bright 
convective  cells are embedded in the brainpattern background.
Their size is very close to the brainpattern. The number and size of
such bright dots increase when reducing the imposed magnetic fields, 
cases D \& E show. The size of these SSC elements 
is smaller than normal granules, inviting comparison 
to the umbral dots observed in
sunspots. This correspondence was indeed suggested by
\cite{Weiss+:fluxsep}; however, the SSC
elements in their experiments only contained
upflows, their upwards directed mass flux being compensated for by a slow
sinking motion in the strongly magnetized bulk fluid. In contrast, we
find strong localized downflows adjacent to the upflows, within the
unmagnetized convective columns. This property makes our SSC
elements a close relative of the {\it convectons} of
\cite{Blanchflower:convecton2D} and  \cite{Blanchflower+:convecton3D}.
So for brevity, in the following we refer to them by this name.

In cases F \& G we recover the typical FXS
phenomenon, first discovered by \cite{Tao+:fluxsep}, where SSC and 
vigorous GRC coexist, with the latter
present in large field free patches (in what follows: {\it inclusions,}
for brevity) embedded in a strongly magnetized fluid displaying SSC in
the form of scattered convectons. Each inclusion includes several
granules. The size of the inclusions is bigger than a normal granule
and smaller than the granules in the deep convective zone. 

Finally, in some regions of the snapshot for case E 
in Fig.~\ref{figcop} displays multiply
segmented convective plumes of a size midway between inclusions and
convectons. The lanes segmenting these elements are cool and coincide
with downflows. This already suggests that our case E represents a
heretofore unknown transitional regime of magnetoconvection between
SSC and fully developed FXS, a finding that will be corroborated below.

\begin{figure}
\centering
\includegraphics[width=\columnwidth]{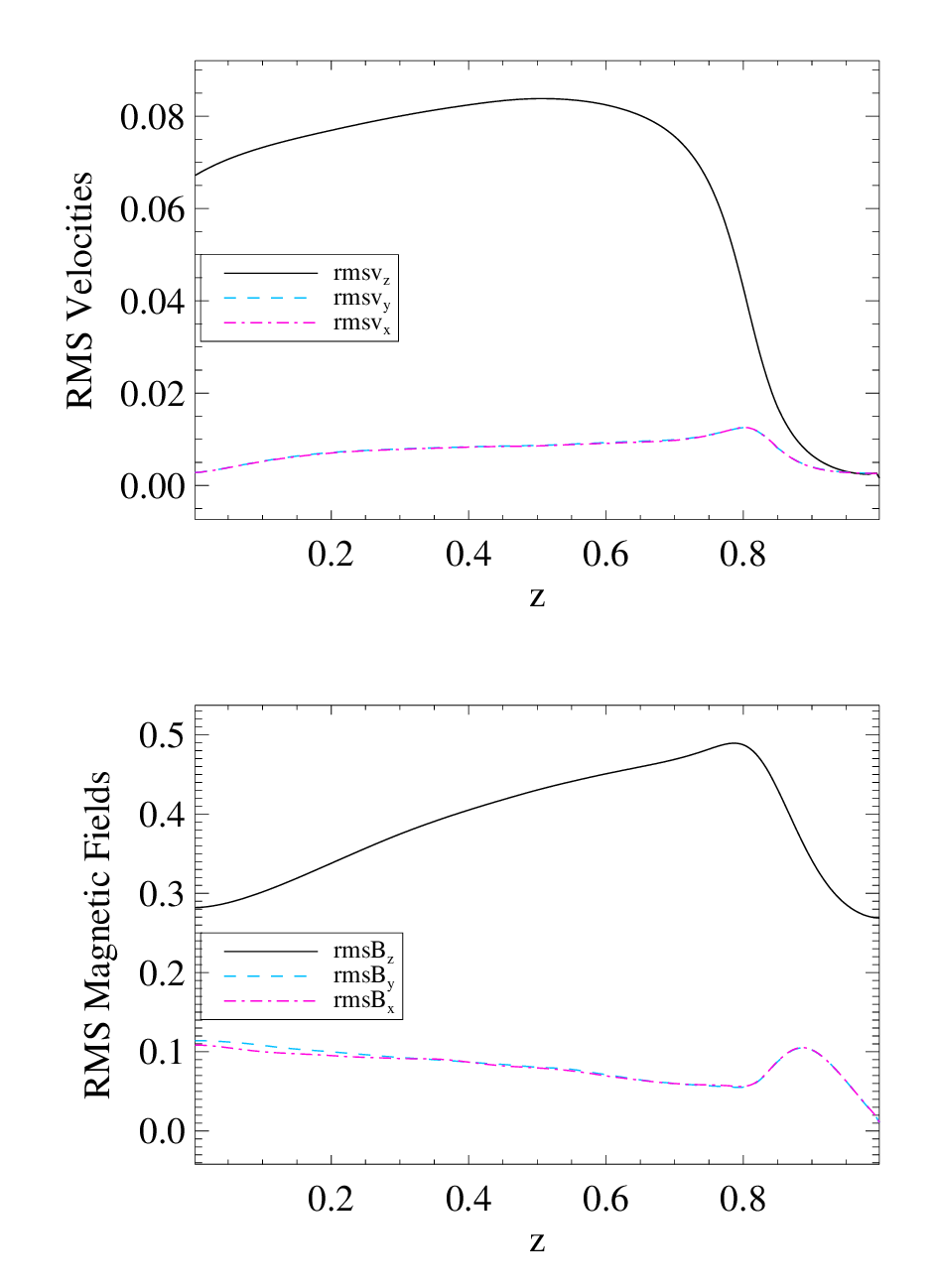}
\caption{Root mean square (rms) velocity and magnetic fields 
from their horizontal means for case F. $z$ is the height from
bottom, in units of $1.72 \rm{Mm}$. Units for velocity and magnetic
field strength are $6.02 \rm{km/s}$ and $2911.55 \rm{G}$, respectively.} 
\label{figrmss}
\end{figure}

Figures \ref{figcosvx} and \ref{figcosmx} present vertical cuts of our
box. As expected, a stronger magnetic field is generally associated with
cooler gas, usually even cooler than in the less magnetized downflows.
In the weak field cases, the downflows show typical von K\'arm\'an
vortices. The magnetic field is amplified and confined by the downward
turbulent flows. In the strong field case, on the other hand, the
turbulent character of the flow is significantly suppressed, field
lines remain close to vertical, and the columnar shape of convectons is
apparent. From Fig.~\ref{figcosvx}, we can see that the maximum of
temperature fluctuation from its horizontal mean occurs near the
convective top. Although for stronger fields, the fluctuation is less
turbulent, its amplitude is higher. 

Both vertical cuts (Figs. \ref{figcosvx} and \ref{figcosmx}) 
and the rms of flow and magnetic fields (Fig.~\ref{figrmss}) show that the
most efficient place where  magnetic fields are amplified by flow motions
is near the top of the convective zone. The enhancement of magnetic field
is not completely coherent with flow motions. The maximum fluctuation
of magnetic fields is higher than the velocity. This kind of enhancement is
significant. For case A, the imposed initial fields are $3\rm{B_e}\sim 
 4455$G, and the amplified strong fields can be $2\rm{B_{scl}}\sim 
 6000$G. In the intermittent case, H, occasionally, the magnetic fields 
confined in pores can also be very strong (bottom panel in Fig.~\ref{figcosmx}).

\begin{figure*}
\centering
\includegraphics[scale=0.9]{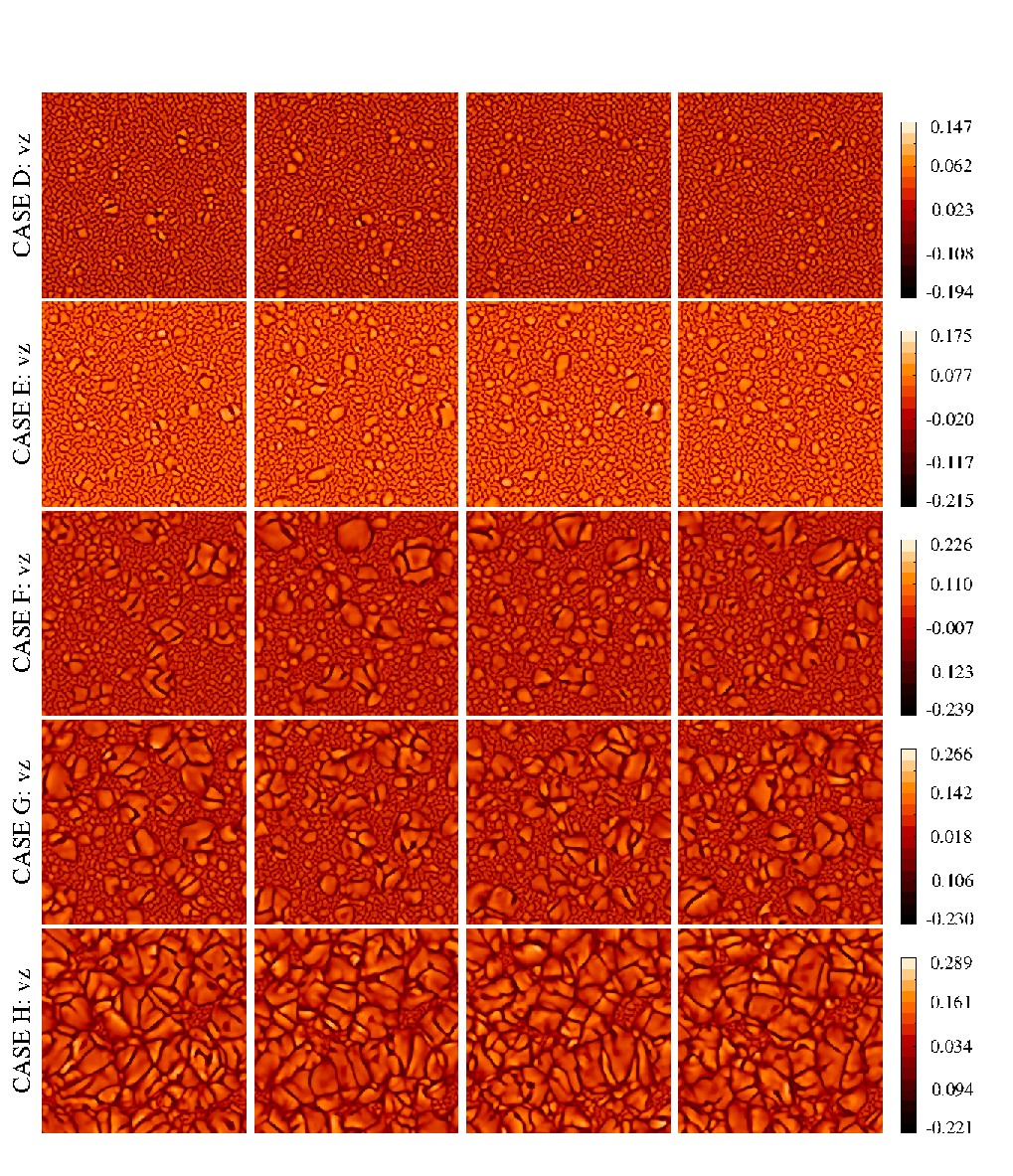}
\caption{
Time evolution of vertical velocity at $z= 0.8$, near the top of the
convective layer. The time intervals between snapshots are
$\sim 71$ minutes for cases D,E,F,G; $\sim 86$ minutes for case H. 
Velocities are given in units of $6.03 \rm{km/s}$. }
\label{figtime}
\end{figure*}

\subsection{Time evolution}

Figure~\ref{figtime} illustrates the time
evolution of vertical velocity near the  top
of the convective layer. The evolution is more clearly seen in the
corresponding animations, which are publicly available at {\tt
http://astro.elte.hu/ \,$\tilde{}$\,kris/magconvmovies/ animations}

Cases C \& D represent SSC. The evolution of convectons 
in these cases is relatively blurred. Case C seems to be a 
critical situation where brightening of some unresolved
tiny convection cells can be seen occasionally. In case D,
the convectons are clearer than in case C. Most of them
are steady, evolving slowly.

The time variation of the convective elements in case E reveals 
a new type of behaviour apparently not seen in previous simulations.
During size-increase phase of some convectons, one or multiple 
downdrafts appear inside them; these downdrafts merge into
a network of internal downflow lanes, giving the convective
elements a segmented appearance reminiscent of granular inclusions. 
(Again, note the analogy with exploding granules,
cf.\citealp{Hirzberger:exploding}.) The inclusions have a lifetime
of $\sim 70$ to $\sim 160$ minutes, depending on their size. 
The single UD between two inclusions lasts around $30$ minutes.
What we see here is essentially convectons developing into inclusions
and back, so it represents a hitherto unseen transitional state
between SSC and FXS.

In the FXS cases, F \& G, we have normal GRC
going on inside the inclusions while small, relatively steady convectons
are occurring in the magnetized fluid in between. These two kinds of 
structures evolve separately at different locations. There seems to be
no more evolution of convectons to or from inclusions. Although there are
smaller granules growing and decaying near the edge of inclusions,
they should not be regarded as convectons.

Compared with granules, pore-like structures in Case H have a much longer
lifetime. In fact, in our simulations, the pores persist ever
since they formed. They are just stretched and moved by convective flows
from one  place to another. The reason  might be that pores
are confined by big granules in the deep convection zone and they have
longer lifetimes than near surface granules. Currently we do not have
enough computational resources to let our models run long enough to
allow obvious deformation and reformation of these deep big granules.

\subsection{Flow statistics}

\begin{figure}
\centering
\includegraphics[width=\columnwidth]{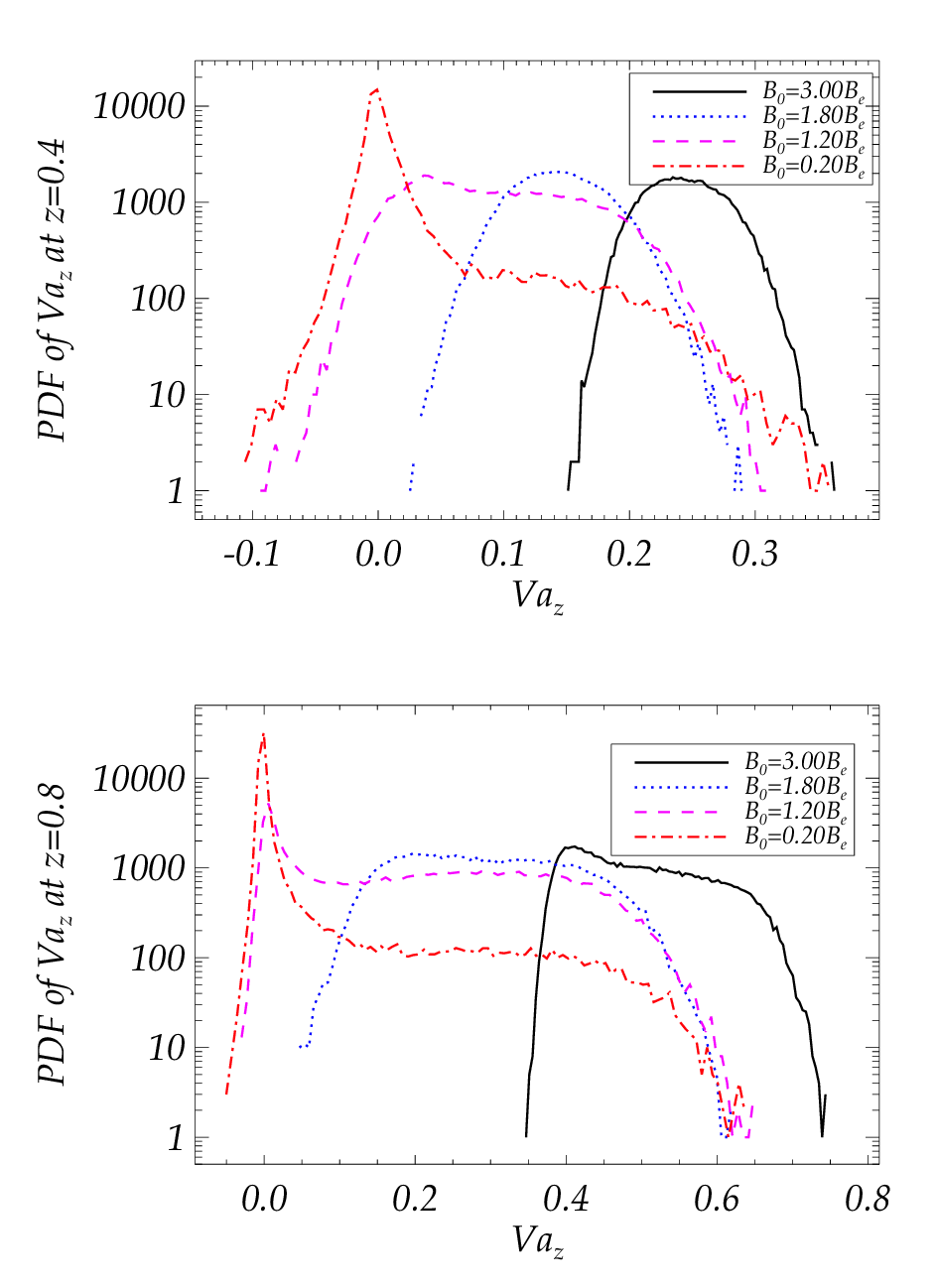}
\caption{
PDFs for the distribution of vertical Alfv\'enic speed,
$Va_z=B_z/\sqrt{4\pi\rho}$, at $z=0.8$ and $z=0.4$ for various
imposed vertical magnetic field strengths as indicated inside the
labels, corresponding to, totally suppressed (black solid),
SSC (blue dotted), FXS (magenta dashed) and GRC
(red dash dot).}
\label{figpdfmag}
\end{figure}

\begin{figure}
\centering
\includegraphics[width=\columnwidth]{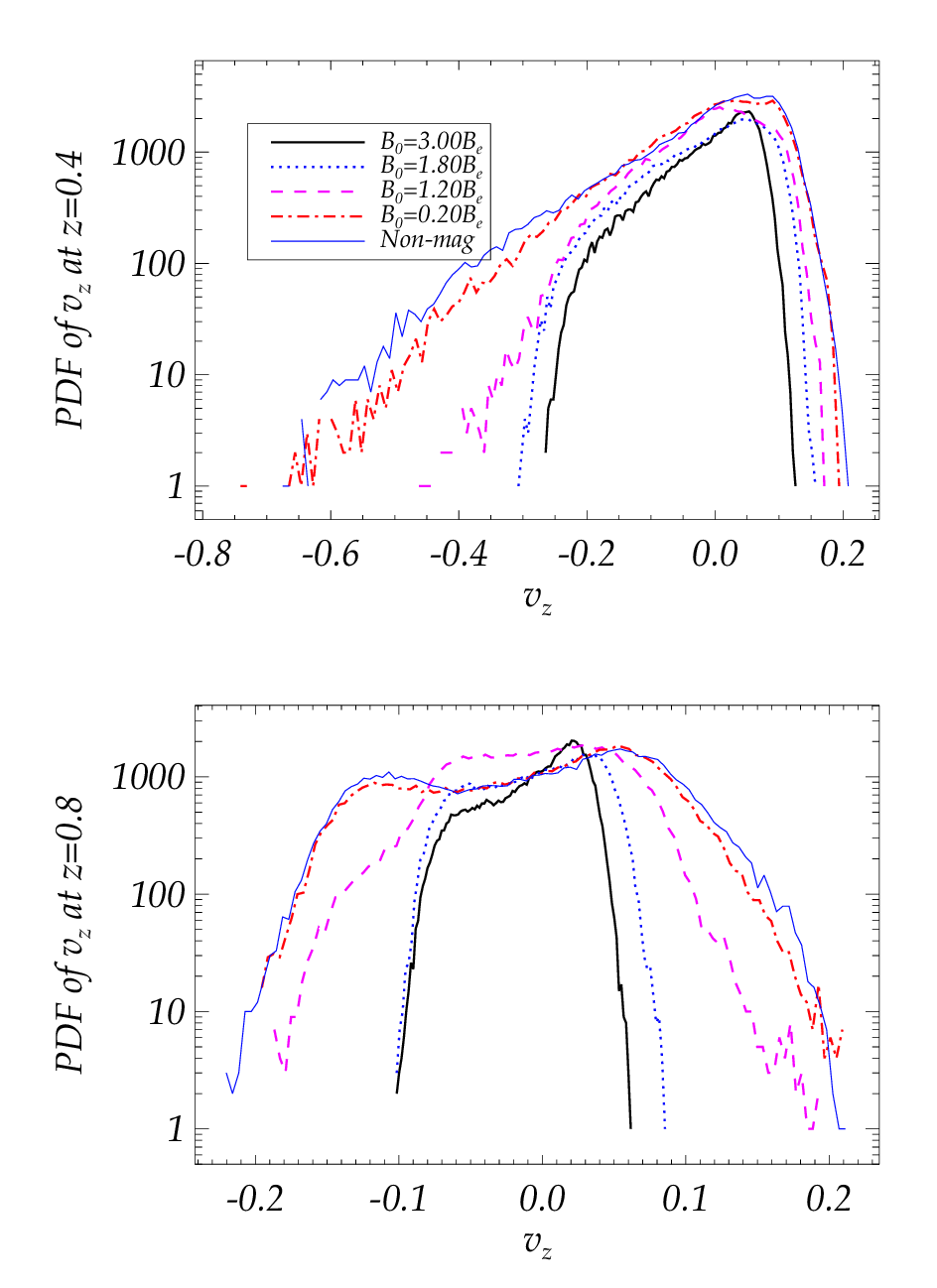}
\caption{
PDFs for the distribution of vertical velocity, i.e.,$v_z$, at $z=0.8$
and $z=0.4$ for various imposed vertical magnetic field strengths
as indicated inside the labels: totally
suppressed field (thick black solid), SSC (blue dotted), FXS
(magenta dashed), and GRC (red dash dot). The blue solid thin
is the PDF of the unmagnetized case.}
\label{figpdfvec}
\end{figure}

For more quantitative analysis of the properties of turbulent
magnetoconvection we plotted the PDFs (probability distribution functions)
of the vertical components of the magnetic field strength (in Alfv\'enic
units) and of the flow velocity in Figs.~\ref{figpdfmag} and
~\ref{figpdfvec}, respectively. Only typical cases were plotted,
i.e., completely suppressed (case A), SSC (case C), 
FXS (case F), and GRC with intermittent fields (case H). 
In Fig.~\ref{figpdfvec} the unmagnetized convection case is also
shown. The data were sampled at two horizontal layers: near the upper
convective boundary ($z\sim 0.8$) and deep inside the convective
part ($z\sim 0.4$).

The shape of PDFs for the distribution of vertical Alfv\'enic speed is
different in the two layers sampled. The width  of PDFs is broader and
sharper in the top of the convection zone than in the middle,  where the
magnetic fields are more intermittent. Our PDFs  differ from those
obtained by \cite{Weiss+:fluxsep} in the strong field region. We do not
have a secondary hump near the maximum. This might be because
our models have transmitting lower boundary and some of
the strong magnetic fields can be transported out of the box freely,
and thus there is no concentration of magnetic fields near the
maximum. 

When the turbulent convection is completely suppressed, the
minimum of Alfv\'enic speed is far from zero, and its PDF
has a narrow width. In such a case, the flow can affect the magnetic
fields but cannot bend them downwards.

For SSC (case C), the PDFs of vertical Alfv\'enic speed are similar 
to case A, but with
a wider width. A minimum of the magnetic fields are approaching zero in
both layers. As we mentioned before, this is highly likely to be a
transition between SSC and a completely suppressed convection case. 
An SSC cell is nearly indistinguishable in this case, 
so we may conclude that
when the imposed vertical field lines are so strong that their
orientation cannot be turned downwards, the convection starts to be
completely suppressed.

For the intermittent field case with GRC (case H),
the forms of the PDF of vertical Alfv\'enic speed 
are similar at different depths. Both have a 
symmetric peak around the origin and a broad slow slope at higher strengths. 
The symmetric peak corresponds to the weakly magnetized, 
vigorously convecting fluid, while the slope is due
to the magnetic flux concentrations. This form of PDF is familiar both
from previous simulations and from quiet Sun observations
(\citealp{dewijn}).

For FXS (case F), the shape of the PDF of vertical Alfv\'enic speed looks
like a combination of GRC and SSC, especially near the convective top.
It has a sharp symmetric peak, a broad flat region, and a quasi
cutoff near the maximum.

The PDFs of the flow velocity sampled in different 
layers show more obvious discrepancies (Fig.~\ref{figpdfvec}).
For GRC with intermittent magnetic field (case H),
the  PDFs of vertical velocity are almost identical to those of
unmagnetized convection. Their change in shape
between the top and the middle of the convecting layer is related to the
well known change in morphology from the observed granular pattern near
the surface to isolated fast downdrafts embedded in a slowly rising,
warm bulk fluid in the bulk of the convective zone. The lack of a 
peak at negative velocities in the top panel may be due to the
low filling factor of the downdrafts and possibly also to a lack of a
characteristic scale among them.

As the imposed magnetic field is increased, the main peak of the PDFs
of the vertical velocity
shifts towards zero. This indicates a general slow sinking of the
magnetized fluid component. An extended shoulder on this peak on the
negative side corresponds to the faster downdrafts localized next to the
upflows. Both this shoulder and the  one on the positive side
representing the upflows are more distinct near the top of the
convecting layer.

\begin{figure}
\centering
\includegraphics[width=\columnwidth]{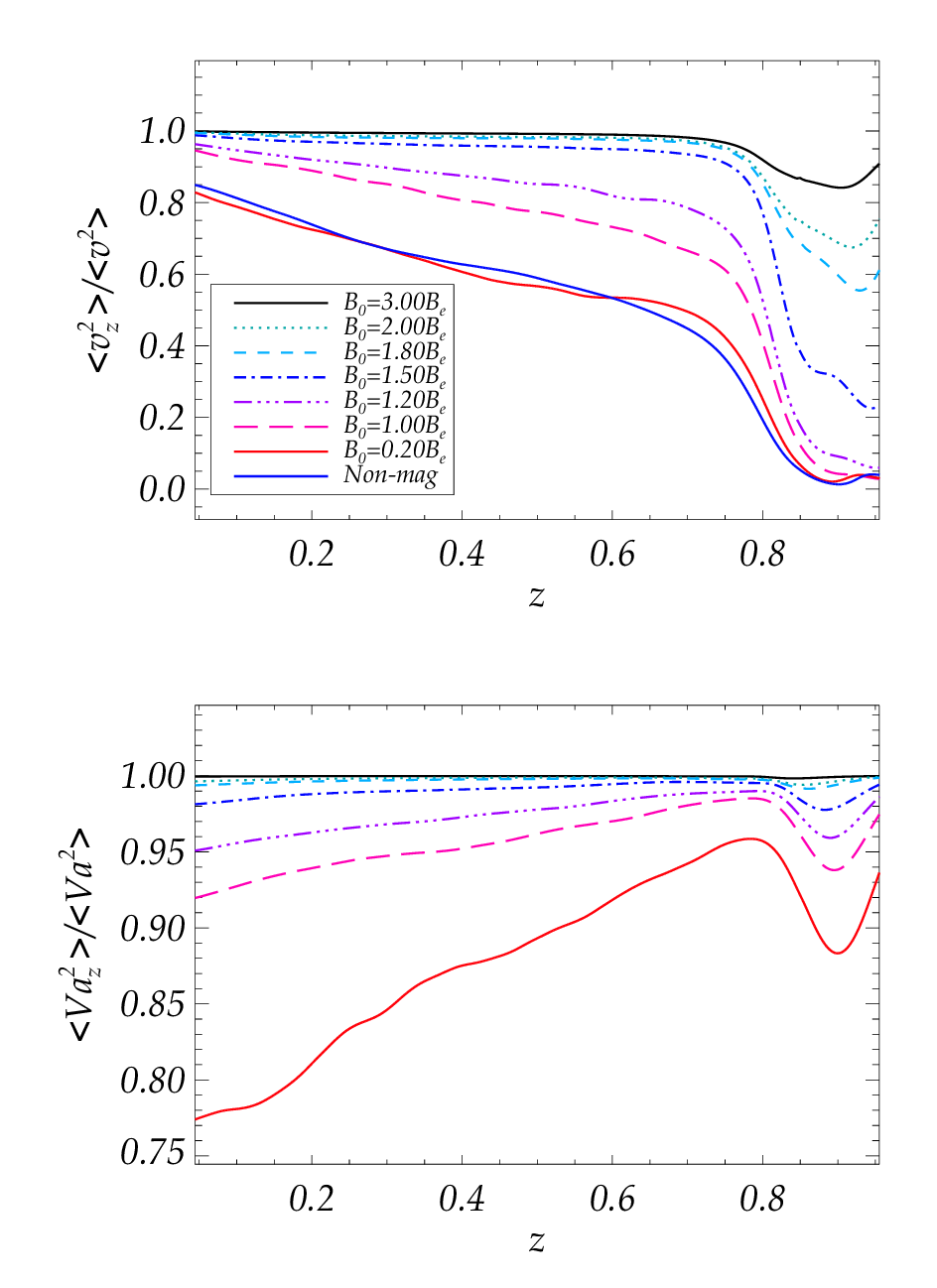}
\caption{Vertical anisotropy of the velocity field (upper panel)
and magnetic field (in Alfv\'enic units, lower panel), for cases A
(thin black solid), B (green dotted), D (cyan dashed), E (blue dash-dot),
F (purple dash--3dot), G (magenta long dashes), and H (red solid).
$z$ is the height from bottom, and $<\ldots>$ represents horizontal
average. }
\label{figcosani1}
\end{figure}

The anisotropy of turbulence is shown in Fig.~\ref{figcosani1}. As
expected, the ratio of vertical component to total field strength
approaches 1 as we increase the imposed magnetic fields (bottom
panel). For the weakest field case in our study, this ratio is around
0.78 near the bottom. The merit of implementing the transmitting lower
boundary condition is evident. The strong anisotropy also occurs near
the top of transition region, at z=0.9. 

In the upper panel, the anisotropy of velocity for unmagnetized convection 
is also plotted. It is clear that the magnetic fields
suppress the flow motions. For the strong magnetic fields, the
velocity field is completely vertical.

\subsection{Comparison with observations}

In the weakly magnetic cases, the morphology, time dependence, and
statistics of our magnetoconvective flow are in good accordance with
observations. This is so  despite the uncertainty of our 
dimensional analysis, which makes it inadequate to directly 
compare the size and lifetime of all kinds of structures with observations.
We attribute this difference mainly to the lack of a
freely radiating photosphere in our experiments, which is known to play
a key role in determining granular scales (\citealp{Rast}). The overall
qualitative and quantitative similarity of the magnetoconvective
structures in our weak field models to observations encourages us to
look for similar observational parallels with the structures seen in our
strong field experiments.

The analysis of high resolution Doppergrams and continuum images of
sunspot umbrae by \cite{Bharti+:DST} show that the umbral dots are often
situated adjacent to downwflows. Localized downflows near UDs have also
been observed by \cite{Ortiz+:SST}. These observed umbral structures are
quite similar in appearance to the convectons surrounded by a patchy
ring of downflows, as seen in our experiments in the SSC domain.

Furthermore, dark lanes are often observed in larger umbral dots. These
UDs rarely have the regular ``coffee bean'' structure suggested by the
simulations of \cite{Rempel+:spotsimu.1D}; instead, dark lanes are often
asymmetrically positioned inside them, and multiple dark lanes are
common, lending a segmented appearance to these UDs. Umbral dots with up
to six dark lanes have been observed by \cite{Bharti+:Hinode.lanes}.
Observations also indicate that the UDs evolve rapidly compared to
sunspots. \cite{Bharti+:Hinode.lanes} show how a UD with multiple dark
lanes changed its appearance over the course of about an hour. (
However, some of the apparent temporal variations in UDs may be due
to seeing effects, cf.~ \citealp{Hamedivafa}.) In the study of
\cite{Schussler+Vogler}, small UDs primarily show a more
symmetric ``bean shape'', while in their Fig. 1 there are also
a few examples of UDs with more complicated dark lanes.
The size of the UD typically depends on how
complicated the substructure is.

These properties of UDs are more reminiscent of the transitional
convective structures seen in our case E. Indeed, case E represents a
vertical magnetic field strength of  $B_0\sim 2.2$~kG in physical units,
considering the uncertainty of dimensional analysis, it is just
in the range expected for sunspot umbrae. We therefore suggest that
conditions in sunspot umbr{\ae} correspond to the FXS-SSC (F/S for
short) transitional regime identified in this paper.

\section{Conclusion}
\label{seccon}

We have constructed a series of numerical experiments of
magnetoconvection in a layer of compressible plasma with an open lower
boundary and bounded by a stable layer from above, using a BGK-MHD
scheme. Our model setup differs from that of previous idealized
magnetoconvection models (\citealp{Weiss+:fluxsep}) in several important
aspects by being closer to realistic solar conditions. Despite this, the
general behaviour of the solutions as a function of the imposed field
strength basically agrees with those models. In particular,
FXS still prevails in a wide parameter range, and the
SSC dominates the energy transport in the strong
field case.

An important difference compared to the results of
\citet{Weiss+:fluxsep} is that in the SSC regime strong, narrow, patchy
downflows appear in a ring around the narrow upflows in our experiments.
Thus, our ``convective columns'' may be more closely related to the
solitary ``convectons'' of \citet{Blanchflower+:convecton3D} than to the
localized plumes in previous magnetoconvection experiments. This
property of our experiments agrees with recent high resolution
observations of sunspot umbrae (\citealp{Bharti+:DST};
\citealp{Bharti+:Hinode.downflows}; \citealp{Ortiz+:SST}).

Furthermore, we have identified a hitherto unknown transitional regime
between FXS and SSC. This F/S regime is very dynamic, with a continuous
evolution of convectons into granular inclusion and back. Intermediate
states of the convective elements are often reminiscent of larger,
multisegmented umbral dots, also known to be subject to dynamic
changes. This suggests that conditions in sunspot umbr{\ae} correspond
to the F/S transitional regime identified in this paper and that UDs
essentially represent convectons and/or convecton/inclusion transitional
phases as seen in our run E.

It should be noted that the realistic numerical simulations of
\cite{Schussler+Vogler} and \cite{Rempel+:spotsimu.1D} suggest a
somewhat different interpretation of the substructure and time
dependence seen in UDs. The dark lanes in these simulations are due to
an optical depth effect and they do not coincide with downflows.
Downflows, however, occur near the end points of the lanes.  High resolution 
observational studies of the correlation and relative
position of downflows and dark lanes may help clarify this issue in
the future.

It should be added that while observations suggest that the properties
of central umbral dots are different from those of peripheric umbral
dots (\citealp{Sobotka+Jurcak:Hinode}), it is hard to draw a sharp line
between the two groups, and the finite tilt of the magnetic field may
play an important role in determining the properties of central UDs as
well. Non-vertical fields, e.g. those due to a fanning out or twisting of the
flux rope, may also be instrumental in inhibiting the appearance of FXS
inside sunspot umbrae ---though this may also be explained
simply by the strong fields in sunspot umbrae or by the large depth (and
consequently low aspect ratio) of the umbra.

These considerations suggest that one possible way to bring idealized
models closer to realistic numerical simulations could be to extend
these results to the case where the imposed magnetic field is tilted.

\begin{acknowledgements}
 We thank the anonymous referee for constructive comments. The
computations were tested and performed on E\"{o}tv\"{o}s University's
756-core, 3.7 Tflop HPC cluster {\it Atlasz} and on the supercomputing
network of the NIIF Hungarian National Supercomputing Centre (project
ID: 1117 fragment). Support by the Hungarian Science Research Fund
(OTKA grants no.\ K81421 and K83133), by the European Commission's 6th
Framework Programme (SOLAIRE Network, MTRN-CT-2006-035484), as well as
by the European Union with the co-financing of the European Social
Fund (grant no.\ T\'AMOP-4.2.1/B- 09/1/KMR-2010-0003) is gratefully
acknowledged. CT acknowledges the financial support from University
of  Helsinki research project `Active Suns'.
\end{acknowledgements}

\bibliographystyle{aa}
\bibliography{magconv}

\end{document}